# Colossal Dielectric Constants in Braced Lattices with Defects

A. P. Ramirez, G. Lawes, V. Butko
*Los Alamos National Laboratories, Los Alamos, NM, 87545*

M. A. Subramanian
*DuPont Central Research and Development Experimental Station*
*Wilmington, Delaware 19880*

C. M. Varma
*Bell Laboratories, Lucent Technologies*
*Murray Hill, NJ 07974*

The dielectric constant of $CaCu_3Ti_4O_{12}$ is measured with a different electrode geometry than previous measurements. Similar colossal values are obtained showing that they are "intrinsic" to the material. We show that in braced structures such as $CaCu_3Ti_4O_{12}$, a small concentration $O(10^{-3})$ of defects, which disrupt the bracing and relax between different equivalent configurations, can lead to colossal dielectric constants above a characterestic temperature. The dielectric constant $e(w,T)$ has a frequency and temperature dependence of the same form as found experimentally.

Dielectric functions $e(w,T)$ with values of the real part of order $10^4$ or higher at low frequencies and with very small imaginary parts (or loss) have been measured recently in some compounds, most notably $CaCu_3Ti_4O_{12}$ [1] [2] for temperatures larger than about 200 K. Such materials have both scientific and technological interest. Systematic and precise calculations have shown that stoichiometric $CaCu_3Ti_4O_{12}$ in its ideal structure is unlikely to have a dielectric constant of more than about $10^2$ [3]. Indeed this is the value found at high frequencies and at low temperatures. We seek an explanation for the extraordinary $e(w,T)$ in terms of a very dilute concentration of a special kind of defect in the perovskite structures.

The first remarkable fact about $CaCu_3Ti_4O_{12}$ is the absence of a structural distortion as a function of temperature and pressure in a cubic structure down to the lowest temperature. This is quite unusual in perovskites. For examples, $CaTiO_3$ has a structural distortion to a tetragonal structure accompanied by a rotation of the $TiO_6$ octahedra at 1400K [4]. Likewise, $SrTiO_3$ has a structural transition at 110 K [5]. $BaTiO_3$ has a ferroelectric transition also accompanied by a rotation of the $TiO_6$ octahedra at 393 K [6]. The distortions in perovskites are empirically correlated with the "tolerance factor", t [7]. For t < 1, as in $CaTiO_3$ [7] the octahedra are rotated, leading to long and short bonds with the corner Ca ions. In $CaCu_3Ti_4O_{12}$ the cubic unit cell of $CaTiO_3$ is quadrupled, with a $Cu^{2+}$ ion replacing three $Ca^{2+}$ ions. Introduction of the smaller $Cu^{2+}$ ions would make the tendency for rotation of octahedra even stronger. But the usual perovskite rules are broken for $CaCu_3Ti_4O_{12}$. It is characteristic of the Jahn-Teller ion $Cu^{2+}$ that it is coordinated by a square of $O^{2-}$ ions and this rule is followed in $CaCu_3Ti_4O_{12}$. Unlike the $Ca^{2+}$ ion, which has a closed electronic shell, the hole in the $Cu^{2+}$ ion has a strong covalent bonding with its square of oxygen neighbors. This leads to a characteristic distance of 1.89Å of the Cu-O bond, observed, for instance, in the "parent" Mott insulating compounds of the high-$T_c$ cuprates. This bond length is also found in $CaCu_3Ti_4O_{12}$. The characteristic energy of distortion of this bond is very high, leading to an optical phonon energy of ~ 0.1eV. Any variation of the $TiO_6$ octahedra necessarily leads to the distortion of the square of oxygens around each Cu in $CaCu_3Ti_4O_{12}$ and is therefore very expensive in energy. The bracing of the perovskite structure by the Cu-O square complex with its very high rigidity is therefore responsible for preserving the cubic structure in $CaCu_3Ti_4O_{12}$.

The other remarkable fact about $CaCu_3Ti_4O_{12}$ is, of course, the extraordinary $\varepsilon(\omega,T)$. In measurements of the ω-dependent capacitance of plates of both densely-packed powder, as well as single crystals, $\varepsilon(\omega,T) > 10^3$ at audio frequencies, and is only weakly varying from above room temperature down to a temperature which varies with frequency indicating relaxational freezing of electric dipoles [2]. This relaxational behavior indicates the dipoles freeze in a relaxor-type state instead of a long-range-ordered state.

Questions have been raised [8] whether the observed $\varepsilon(\omega,T)$ is *intrinsic* to the material or whether it is due to charged states induced at the interface between a sample with a high concentration of charged defects and the metal-electrodes used to measure $\varepsilon(\omega,T)$. This issue can be settled by eliminating the direct metal contact of the sample by interposing other insulators of known properties. We have re-measured the dielectric constant of $CaCu_3Ti_4O_{12}$ with a different electrode geometry than what has been reported previously. We polished parallel faces of a $CaCu_3Ti_4O_{12}$ polycrystalline sample, and deposited 1000 Å of $Al_2O_3$ on these surfaces, followed by 1000 Å of Au to make the metal electrode. The presence of the insulating aluminum oxide layer precludes the possibility of charge accumulating at the interface between the sample and electrode. The capacitance was measured using an Agilent 4284A LCR meter in a commercial cryostat.

The results of this modification of the experimental setup is shown in fig. 1. We note that the dielectric constant in the high-temperature region is slightly less than previously reported with direct contact between metal electrode and $CaCu_3Ti_4O_{12}$. For several samples prepared in this manner, however, the bulk $\varepsilon(\omega,T)$ is about 50 times greater than that predicted using density functional theory [3]. Similar results were also found with the single crystal used in a previous study [9]. It is therefore reasonable to attribute the observed large dielectric constant to a bulk effect, as originally proposed.

We seek an explanation for the extraordinary $\varepsilon(\omega,T)$ of $CaCu_3Ti_4O_{12}$ in terms of isolated defects in the ideal structure. These defects may be Cu vacancies or interchange of Ca and Cu sites such that locally the rigid square Cu-O complex is disrupted. Because the bracing is locally absent, the defective regions have a tendency to undergo distortions characteristic of the perovskite structure discussed above. But since they are isolated, the co-operative effects are absent at low densities. The defect cells will then, in general, relax between alternate equivalent configurations preserving, on average, the cubic structure. The relaxation occurs at a (classical) rate $\gamma$ determined by the energy barrier between alternate configurations and the temperature:

$$\gamma = \gamma_0 \exp(-\Delta/T). \qquad (1)$$

One expects $\gamma_0$ to range between $O(10^{10}-10^{12})$ Hz, depending on the effective mass of the defect and $\Delta$ to be of $O(0.1)$ eV. Such defects were hypothesized to explain the "central-peak" dynamics near structural transitions including the 110 K transition in $SrTiO_3$ [10]. In the latter, direct evidence for their presence was found through EPR experiments by Muller et al. [11]. The defects will in general be polarizable with a *local* polarizability,

$$\chi_i(\omega,T) = \frac{\mu_i \gamma}{-i\omega + \gamma} \qquad (2)$$

where $\mu_i$ is the static polarizability of the isolated defect, i.e. dipole moment per unit applied field and has dimension of inverse density.

To calculate the dielectric function of the lattice with defect cells, we adopt a mean-field approximation in which we average over the impurity configurations by introducing a two component displacement or fluctuating dipole in each unit cell, one for the pure component and the other due to the defective cells. Details of this impurity averaging procedure, which is similar to the "Average T-matrix Approximation" (ATA) in the theory of random systems [12] may be found in Ref. [10]. We also assume that the dielelectric properties are due to localized excitations, so that the *local-field effects* can be well handled in the Clausius-Mosotti approximation.

The polarization in each unit cell is the vector:

$$\mathbf{p} = \begin{pmatrix} p_p \\ p_i \end{pmatrix} \qquad (3)$$

where $p_p$ is the pure component and $p_i$ is the defect component. $\mathbf{p}$ obeys

$$\mathbf{p} = \Xi(\omega,T)\mathbf{E}_0(\omega) = \Xi^0(\omega,T)\mathbf{E}(\omega). \qquad (4)$$

Here $\mathbf{E}_0$ is the applied field and $\mathbf{E}$ is the *internal* field. $\Xi^0$ is the diagonal matrix of the response functions without local field effects made by the response of the pure component $\chi_p$ and the response of the impure component $\chi_i$. $\Xi$, the response including the local field effects is constructed from considerations similar to those in the Clausius-Mosotti approximation.

If we were dealing with a single component polarization instead of Eq. 3, the response would be given in the Clausius-Mosotti approximation by $\Xi^{-1}(\omega,T) = \chi^{-1} - (4\pi/3)\rho$. The generalization to the two-component system of Eq. 3 is that the inverse of the $\Xi$ matrix is given by

$$\Xi(\omega,T)^{-1} = \begin{pmatrix} \chi_p^{-1} - (4\pi/3)(1-c)\rho & -(4\pi/3)c\rho \\ -(4\pi/3)(1-c)\rho & \chi_i^{-1} - (4\pi/3)c\rho \end{pmatrix} \qquad (5)$$

where $\rho$ is the density of dipoles in the pure material. The averaged polarizability is then given by

$$\langle \Xi(\boldsymbol{w}, \mathrm{T}) \rangle = (1-c)(\hat{\boldsymbol{c}}_{pp} + \hat{\boldsymbol{c}}_{pi}) + c(\hat{\boldsymbol{c}}_{ii} + \hat{\boldsymbol{c}}_{ip}) \qquad (6)$$

Here $\hat{\boldsymbol{c}}_{ab}$ is the a-b component of $\Xi$ obtained from Eq. (7) after inversion.

We next assume, as is reasonable for considering the low-frequency region in materials such as CaCu$_3$Ti$_4$O$_{12}$ that $\chi_p$ is independent of frequency and purely real. This means that the oscillators responsible for the low-frequency dielectric function of the hypothetically pure material have absorptions at frequencies much higher than our range of interest. Then with the dielectric constant $\varepsilon_0$ of such a material $\gg 1$ and $c \ll 1$, the leading contribution to the average dielectric function is

$$\langle \varepsilon(\omega, \mathrm{T}) \rangle \approx \frac{\varepsilon_0}{1 - \frac{4\pi}{3}\rho c \varepsilon_0 \frac{\mu\gamma}{-i\omega + \gamma}}. \qquad (9)$$

We may look at the limiting form of this expression to see the observed phenomena emerging and its physical basis. For $\omega \gg (4\pi/3)c\rho\chi_0\gamma(\mathrm{T})$, i.e. at low enough temperatures when the impurity cells are frozen, $\qquad (10)$

$$\langle \varepsilon(\omega, \mathrm{T}) \rangle \approx \varepsilon_0 (1 - i\gamma(\mathrm{T})/\omega).$$

In the opposite limit when temperature is high enough that $\gamma(\mathrm{T}) \approx \gamma_0$ and $\omega \ll \gamma_0$

$$\langle \varepsilon(\omega, \mathrm{T}) \rangle \approx \frac{\varepsilon_0 (1 - i\omega/\gamma(\mathrm{T}))}{1 - (4\pi/3)c\rho\varepsilon_0\mu} \qquad (11)$$

In this case the dielectric constant is enhanced by a factor
$$\Psi \approx (1 - (4\pi/3)c\rho\varepsilon_0\mu)^{-1}. \qquad (12)$$

due to the riding of the local-fields on the relaxing cells on the large background dielectric constant $\varepsilon_0$ discussed elsewhere [3, 13]. In effect the effective charge of the defect cells has gone up by a factor $\varepsilon \approx 100$ for CaCu$_3$Ti$_4$O$_{12}$.

The limitations of the approximations in which Eq. (9) is derived should be kept in mind. The ATA averaging approximation does not properly consider the variations around each defect cell properly. Defect cells with a large dipole moment or smaller $\Delta$ than the average will freeze out as the temperature is lowered before the average and lost to the dynamical processes under consideration. A better treatment will therefore have an effective concentration, c, for dynamical processes decreasing with temperature. For large enough bare defect concentration a co-operative transition to a glassy state must occur and the dynamical enhancements to the dielectric functions considered here are the lost. This is subject to experimental test by purposely damaging the material. Another experimental test is to monitor the diffuse scattering due to relaxing cells by, for example, neutron scattering. Large diffuse scattering should be seen in the regime of large dielectric constants with freezing at lower temperatures.

The complete expression, Eq. (9) is plotted and compared with experiments in Figs. 1 and 2. The parameters employed are shown with the dimensionless parameter $a \equiv (4\pi/3)\rho c \varepsilon_0 \mu$. We note that with the reasonable value for the polarizability of the defect such that $\rho\mu \cong 1$ and $\varepsilon_0 \approx 10^2$, only $c \sim O(3 \times 10^3)$ is required. The activation energy of the defect is $O(10^3 K)$, which is also a reasonable value. The general shape of the real and imaginary part of the dielectric constant with respect to both the frequency and the temperature dependence is well reproduced. Some discrepancies are evident which are worth remarking. First, there is evidence for another plateau in the real part of the dielectric constant at a higher temperature. This could be reproduced by the theory if a concentration of defects with a higher activation energy were added. This strength of the extra plateau varies from sample to sample and is often found not at all. A common discrepancy between the experimental results and the theory is that the experimental curves have a smoother crossover from the low temperature regime to the high temperature plateau. This discrepancy is due to the simplicity of our model. A treatment with a distribution in properties of defects as discussed above will clearly change the results in the correct direction. Rather than use a more elaborate theoretical treatment to fit the data, it is perhaps more useful to test the model with more experiments as suggested above.

**Figure Captions**

1. The real part of the dielectric constant, $\varepsilon(\omega, T)$, as a function of temperature, for three different frequencies, $\omega/2\pi$. The solid lines are fits to equation 11, using the parameters $\gamma_0 = 8 \times 10^8$ Hz, $E_a/k_B = 800$K, and $a = 0.9805$.

2. The imaginary part of the dielectric constant, $\varepsilon(\omega, T)$ as a function of temperature, for two different frequencies, $\omega/2\pi$. The solid and dashed lines are fits to the data using Eq. 11 and the same parameters used in fig. 1.

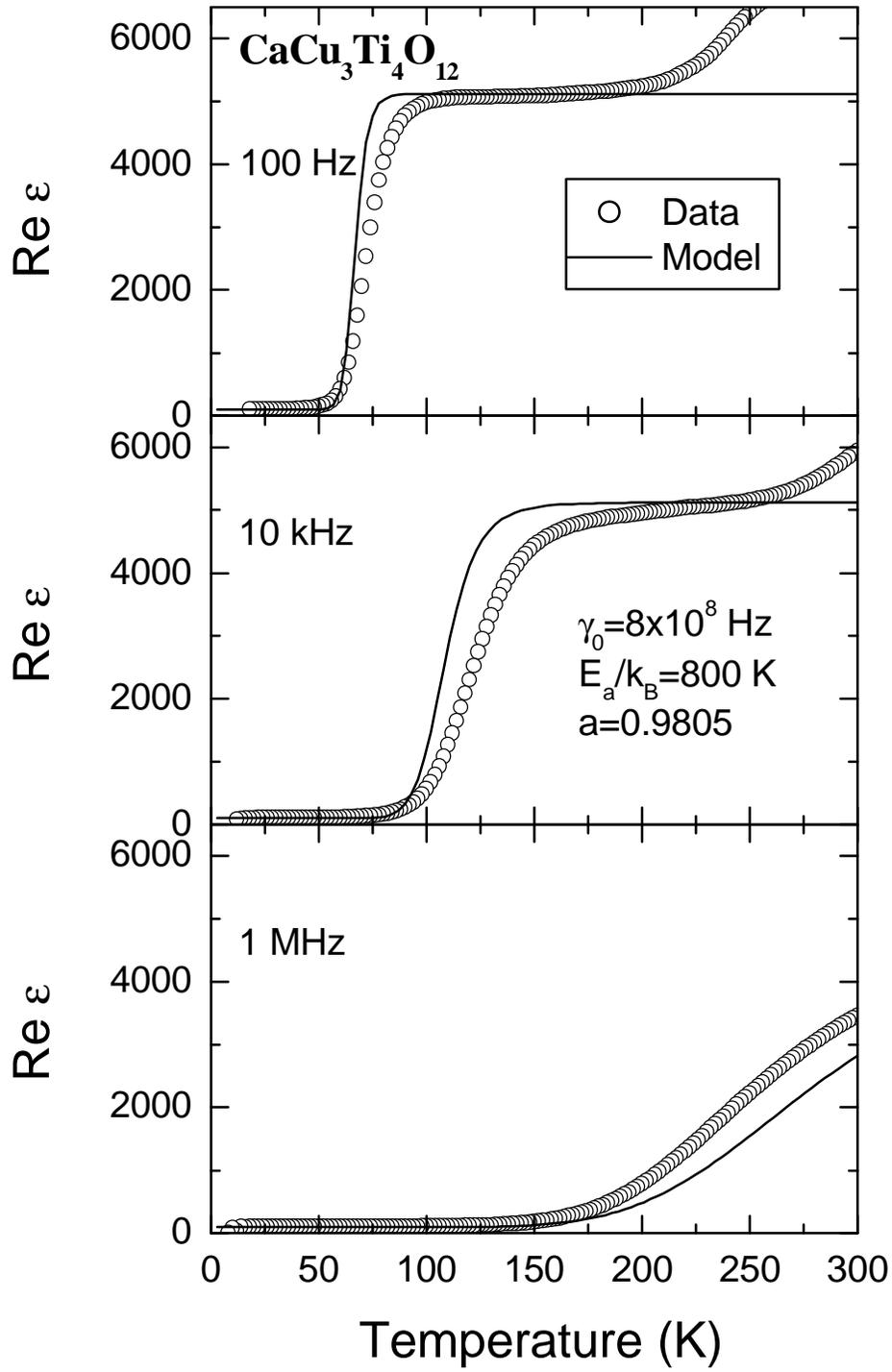

Fig. 1

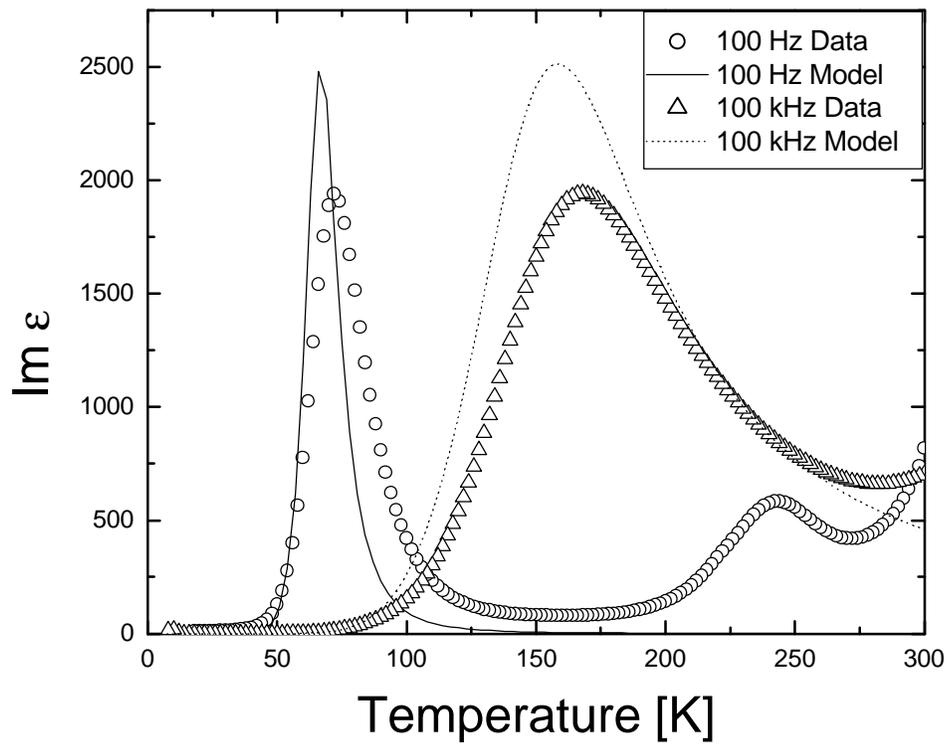

Fig. 2